\documentclass[
 reprint,
 amsmath,amssymb,superscriptaddress,
 aps,
 prl,
groupedaddress
]{revtex4-1}

\usepackage{graphicx}
\usepackage{dcolumn}
\usepackage{bm}
\usepackage{color}
\usepackage{ulem}
\usepackage{float}
\usepackage{textcomp}
\restylefloat{table}
\usepackage{verbatim}
\usepackage{epstopdf}

\begin{document}

\title{Search for Composite Fermions at Filling Factor 5/2: Role of Landau Level and Subband Index}
\date{\today}

\author{M. A.\ Mueed}
\author{D.\ Kamburov}
\author{Md.\ Shafayat Hossain}
\author{L. N.\ Pfeiffer} 
\author{K. W.\ West}
\author{K. W.\ Baldwin}
\author{M.\ Shayegan}
\affiliation{Department of Electrical Engineering, Princeton University, Princeton, New Jersey 08544, USA}

\begin{abstract}
The pairing of composite fermions (CFs), electron-flux quasi-particles, is commonly proposed to explain the even-denominator fractional quantum Hall state observed at $\nu=5/2$ in the first excited ($N=1$) Landau level (LL) of a two-dimensional electron system (2DES). While well-established to exist in the lowest ($N=0$) LL, much is unknown about CFs in the $N=1$ LL. Here we carry out geometric resonance measurements to detect CFs at $\nu=5/2$ by subjecting the 2DES to a one-dimensional density modulation. Our data, taken at a temperature of 0.3 K, reveal no geometric resonances for CFs in the $N=1$ LL. In stark contrast, we observe clear signatures of such resonances when $\nu=5/2$ is placed in the $N=0$ LL of the anti-symmetric subband by varying the 2DES width. This finding implies that the CFs' mean-free-path is significantly smaller in the $N=1$ LL compared to the $N=0$ LL. Our additional data as a function of in-plane magnetic field highlight the role of subband index and establish that CFs at $\nu=5/2$ in the $N=0$ LL are more anisotropic in the symmetric subband than in the anti-symmetric subband.   
\end{abstract} 

\maketitle

\begin{figure*}
\includegraphics[width=.98\textwidth]{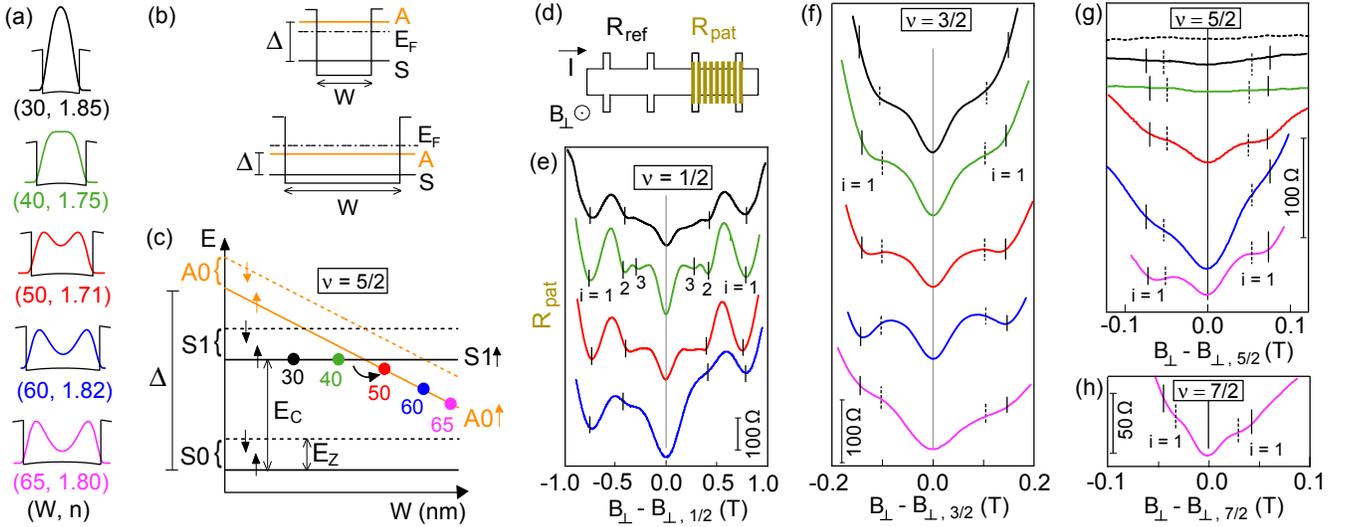}
\caption{\label{fig:Fig1} (color online)  (a) Self-consistently calculated charge distributions (color coded) at $B=0$ for different sample parameters $(W,n)$ in the units of (nm, $10^{11}$ cm$^{-2}$). (b) Reduction of $\Delta$ via increasing $W$ (see text). (c) The evolution of the lowest two LLs ($S0$ and $S1$) of the $S$ and the lowest LL ($A0$) of the $A$ subband as a function of $W$ at $\nu=5/2$. Solid and dashed lines refer to the spin-up ($\uparrow$) and spin-down ($\downarrow$) levels. All the LLs are drawn with respect to the lowest energy level $S0$$\uparrow$ which is shown as constant. In a simple picture, the cyclotron energy ($E_{C}$) and the Zeeman energy ($E_{Z}$), which is much smaller than $E_C$ in GaAs, do not vary with $W$ and remain fixed at their $\nu=5/2$ values. 
%
However, $\Delta$ decreases with increasing $W$, inducing several LL crossings. The relevant crossing for $\nu=5/2$ is between $S1$$\uparrow$ and $A0$$\uparrow$: $\nu=5/2$ resides in the $S1$$\uparrow$ LL if $E_C<\Delta$, but moves to the $A0$$\uparrow$ LL when $E_C>\Delta$. Based on the values of $E_C$ and $\Delta$ (see text), we mark with solid circles the $\nu=5/2$ positions for all our samples. (d) Sample schematic. The brown lines represent the surface superlattice. Part of the sample is intentionally left unpatterned for reference. The patterned and unpatterned regions each have a length of 100 $\mu$m and width of 50 $\mu$m.
(e)-(h) Magnetoresistance traces, taken at $T=0.3$ K, for different samples near $\nu=1/2$, 3/2, 5/2, and 7/2 showing GR features of CFs when they reside in an $N=0$ LL. The solid and dotted lines mark the expected positions of the primary ($i=1$) GR feature of spin-polarized and unpolarized CFs, respectively. In (e) we also mark the expected positions of the higher-order GR features ($i=2$ and 3). All the traces are for the patterned regions, except for the top (dotted black) trace in (g) which is for the unpatterned region of the $W=30$ nm sample. Traces in all panels are shifted vertically for clarity, and follow the color code of (a). Also, the magnitude of $B_{\perp}$ at a given filling factor $\nu$ is equal to $(h/e)(n/\nu)$.}
\end{figure*}

Ever since its discovery in clean two dimensional electron systems (2DESs), the fractional quantum Hall state (FQHS) at $\nu=5/2$ in the first excited ($N=1$) Landau level (LL) has been an enigmatic topic in condensed matter physics \cite{Jain.2007, Shayegan.Flatland.Review.2006,Willett.PRL.1987,Nayak.RMP.2008}. The interest has been fueled by the possibility of this state's non-Abelian nature and implementation in topological quantum computation \cite{Nayak.RMP.2008}. The most promising explanation for the $\nu=5/2$ FQHS involves pairing physics of composite fermions (CFs) \cite{Haldane.PRL.1988,Moore.NuPhy.1991,Greiter.PRL.1991,Morf.PRL.1998,Rezayi.PRL.2000,Read.Physica.2001}, exotic quasi-particles that are products of electrons and flux quanta \cite{Jain.2007,Jain.PRL.1989,Halperin.PRB.1993}.
These studies assume that CFs occupy a weakly interacting Fermi sea at high temperatures at $\nu=5/2$ but pair up to form a FQHS at low temperatures as they become more interacting. Now, while the existence of CFs in the lowest ($N=0$) LL, i.e, near $\nu=1/2$ and 3/2, is well-established theoretically and experimentally through geometric resonance (GR) studies \cite{Halperin.PRB.1993,Willett.PRL.1993, Kang.PRL.1993,Vladimir.PRL.1994, Smet.PRL.1996, Smet.PRB.1997, Smet.PRL.1998, Smet.PRL.1999, Willett.PRL.1999, Kamburov.PRL.2014,Willett.AP.1997,Willett.PRL.1997,Endo.PRB.2001,Kamburov1.PRB.2014,Kamburov.PRL.2012,Kamburov.PRB.2014,Kamburov.PRL.2013,Mirlin.PRL.1998, Oppen.PRL.1998,Zwerschke.PRL.1999}, such conclusive observation remains elusive for $\nu=5/2$ although there are hints from surface acoustic wave (SAW) experiments \cite{Willett.PRL.2002,Willett.RPP.2013}. Here we carry out GR measurements on several 2DESs confined to GaAs quantum wells (QWs) and subjected to a one-dimensional density modulation. By varying the QW width (Figs. \ref{fig:Fig1}(a) and (b)), we tune the Fermi level at $\nu=5/2$ between the $N=1$ LL of the symmetric subband and the $N=0$ LL of the anti-symmetric subband (see Fig. \ref{fig:Fig1}(c)). The data reveal an absence of GR features for $\nu=5/2$ when it resides in the $N=1$ LL. In marked contrast, we observe pronounced GR features when $\nu=5/2$ forms in the $N=0$ LL.

Before a detailed presentation, we briefly discuss CFs and their detection through GR measurements \cite{Jain.2007,Halperin.PRB.1993}. At $\nu=1/2$, each CF is modeled as one electron bound to two magnetic flux quanta and is effectively shielded from the applied perpendicular magnetic field ($B_{\perp}$) due to the flux attachment. As a result, CFs behave as if $B_{\perp}=0$ at $\nu=1/2$, thus leading to the formation of a Fermi gas similar to electrons at $B_{\perp}=0$. Away from $\nu=1/2$, however, CFs feel an effective magnetic field $B_{\perp}^{*}=B_{\perp}-B_{\perp,1/2}$ \cite{footnote1}, where $B_{\perp, 1/2}$ is the field at $\nu=1/2$. When subjected to $B_{\perp}^{*}$, they move in a cyclotron orbit whose diameter ($2R_{C}^{*}=2\hbar k_F^{*}/eB_{\perp}^{*}$) is determined by the CF Fermi wave vector $k_F^{*}=(4\pi n_{CF}^{*})^{1/2}$ where $n_{CF}^{*}$ is the CF density. CFs also form near $\nu=3/2$ (= 1+1/2), the half-filling of the opposite-spin $N=0$ LL. However, $n_{CF}^{*}$ at $\nu=3/2$ is 1/3 of the carrier density ($n$) since the lower LL is fully occupied and effectively inert. Similarly, $n_{CF}^{*}$ should be $n/5$ if CFs exist at $\nu=5/2$ (= 2+1/2).

GR measurements are typically implemented using periodic perturbation techniques such as applying SAWs \cite{Willett.PRL.1993,Willett.PRL.1997,Willett.AP.1997} or imposing a one-dimensional density modulation \cite{Willett.PRL.1999, Smet.PRB.1997, Smet.PRL.1998, Mirlin.PRL.1998, Oppen.PRL.1998, Smet.PRL.1999, Zwerschke.PRL.1999, Kamburov.PRL.2014,Endo.PRB.2001,Kamburov1.PRB.2014,Kamburov.PRB.2014,Kamburov.PRL.2012,Kamburov.PRL.2013}. In the latter method, if CFs near $\nu=1/2$ or 3/2 can complete a cyclotron orbit without scattering, then a GR takes place between
their orbit diameter and the period ($a$) of the perturbation. When the condition for CFs' GR, i.e., $2R_{c}^{*}/a=i+1/4$ ($i=1,2,3,...$) \cite{Smet.PRB.1997, Smet.PRL.1998, Mirlin.PRL.1998, Oppen.PRL.1998, Smet.PRL.1999, Zwerschke.PRL.1999,Endo.PRB.2001, Willett.PRL.1999,Kamburov.PRL.2014,Kamburov.PRL.2012,Kamburov.PRL.2013,Kamburov.PRB.2014,Kamburov1.PRB.2014} is satisfied, the resistance near $\nu=1/2$ or 3/2 exhibits a minimum. The $B_{\perp}^{*}$-position of the GR features directly measures $k_F^{*}$, thus conclusively proving the existence of a CF Fermi gas near $\nu=1/2$ or 3/2. 

Our samples consist of a single GaAs QW buried 190 nm underneath the surface, with 150-nm-wide Al$_{0.24}$Ga$_{0.76}$As barrier layers on each side. The QW width ($W$) and $n$ for these samples vary from $30-65$ nm and $1.71-1.85\times10^{11}$ cm$^{-2}$ (Fig. \ref{fig:Fig1}(a)). The corresponding charge-distributions, calculated self-consistently at $B=0$, are shown in Fig. \ref{fig:Fig1}(a) in a color code, also used for all the magnetoresistance traces in Fig. \ref{fig:Fig1}. For similar $n$, the charge-distribution profile clearly becomes more bilayer-like as $W$ is increased. This is because the anti-symmetric ($A$) subband, whose wave function possesses a node along the growth (out-of-plane) direction, gets progressively more occupied since its energy separation ($\Delta$) from the symmetric ($S$) subband decreases for large $W$ (see Fig. \ref{fig:Fig1}(b)). The consequence of increasing $W$ is illustrated in Fig. \ref{fig:Fig1}(c) where we outline the evolution of the $N=0$ and 1 LLs from the $S$, and the $N=0$ LLs from the $A$ subband. As shown, for these spin-resolved (split by the Zeeman energy $E_Z$) LLs, $\Delta$ diminishes when $W$ is increased but the LL separation (cyclotron energy $E_C$) remains unchanged. For our samples, $E_C$ at $\nu=5/2$ is between $55-60$ K while $\Delta$ changes from 144 K to 13 K as $W$ varies from 30 to 65 nm. Because $\Delta>E_C$ for $W=30$ and 40 nm, $\nu=5/2$ resides in the $N=1$ LL, but moves to the $N=0$ LL of the $A$ subband when $\Delta<E_C$ for larger $W$ (= 50, 60, and 65 nm). We schematically show this evolution in Fig. \ref{fig:Fig1}(c) by marking the positions of $\nu=5/2$ in the corresponding LLs with solid circles, color coded similar to Fig. \ref{fig:Fig1}(a)).

The above evolution as a function of $W$ allows us to study the CF Fermi gas at a fixed $\nu$ in two different LLs. It is worth noting that such a LL transition, i.e., between $N=1$ and 0, is not possible for $\nu=3/2$ or 1/2 in GaAs 2DESs. 
To probe the CFs via the GR technique, we partially pattern the surface of our samples, standard Hall bars (see Fig. \ref{fig:Fig1}(d)), with a strain-inducing superlattice. Because of the piezo-electric effect in GaAs, this pattern of period $a=200$ nm, made of negative electron-beam resist, causes a density modulation of the same period in the 2DES \cite{Endo.PRB.2005,Cusco.Surf.1994,Kamburov.PRB.2012,Endo.PRB.2000,Skuras.APL.1997,Long.PRB.1999,Endo.PRB.2005}. The low-temperature mobilities of our samples, $\simeq1\times10^{7}$ cm$^{2}/$Vs, are very high and therefore favorable to ballistic transport of CFs, a necessary condition for GR phenomena. Measurements were done in $^3$He refrigerators at 0.3 K via passing current perpendicular to the density modulation (Fig. \ref{fig:Fig1}(d)).

We first address the GR of CFs in the $N=0$ LL, i.e., for $\nu=1/2$ and 3/2. Figures \ref{fig:Fig1}(e) and (f) present a series of magnetoresistance traces near $\nu=1/2$ and 3/2 plotted as a function of $B_{\perp}-B_{\perp,1/2}$ and $B_{\perp}-B_{\perp,3/2}$, respectively. For all cases, we observe a marked resistance minimum at $\nu=1/2$ or 3/2, followed by additional resistance minima or shoulders on the flanks; these features are characteristic of the GR phenomenon. With solid and dotted lines, we mark the expected field positions of the $i=1$ GR feature assuming spin-polarized and spin-unpolarized CFs, respectively. Note that $n_{CF}^{*}$ is effectively halved for spin-unpolarized CFs. The GR features (Fig. \ref{fig:Fig1}(e)) near $\nu=1/2$ are consistent with fully spin-polarized CFs for all $W$ \cite{footnote1_1}. For $\nu=3/2$, however, CFs become more spin-polarized with increasing $W$ \cite{Kamburov1.PRB.2014,Liu.PRB.2014}. Their GR features, resembling resistance shoulders (Fig. \ref{fig:Fig1}(f)), are closer to the dotted lines for $W=30$ and 40 nm. For larger $W$, the GR features agree better with the solid lines and are also more pronounced \cite{footnote1_2}. 

Next, we present the magnetoresistance traces near $\nu=5/2$ in Fig. \ref{fig:Fig1}(g) as a function of $B_{\perp}-B_{\perp,5/2}$ for different $W$.
For reference, we also include the unpatterned region's trace (dotted black) for the $W=30$ nm sample. This trace shows an almost flat resistance profile across $\nu=5/2$ and no sign of GR features, as expected. The patterned region's traces for $W=30$ (black) and 40 nm (green), when $\nu=5/2$ is in the $S1$$\uparrow$ LL (Fig. \ref{fig:Fig1}(c)), are also similarly featureless. Absence of GR features was also observed for two other 2DES samples where $\nu=5/2$ forms in the $N=1$ LL ($W=30$ nm and $n\simeq1.5$ and $\simeq2.9\times10^{11}$ cm$^{-2}$). In contrast, when $\nu=5/2$ resides in the $A0$$\uparrow$ LL for $W\geq50$ nm (Fig. \ref{fig:Fig1}(c)), we observe GR features in Fig. \ref{fig:Fig1}(g) similar to the $\nu=1/2$ and 3/2 cases. Moreover, the solid lines (expected $i=1$ GR feature position for spin-polarized case) agree well with the resistance shoulders or minima on both sides of $\nu=5/2$, suggesting that these features are indeed the GR of fully spin-polarized CFs. In light of Fig. \ref{fig:Fig1}(c), we conclude that GR features are seen in the $N=0$ ($A0$$\uparrow$) LL ($W\geq50$ nm) but \textit{not} in the $N=1$ ($S1$$\uparrow$) LL ($W\leq40$ nm). Note that the contrast implied by our data corroborates a previous observation: when $\nu=5/2$ is placed in an $N=0$ LL in a wide QW, its nearby FQHSs appear in a simple, odd-denominator series \cite{Shabani.PRL.2010,Liu.PRB.2011}, similar to the usual FQHS series seen flanking $\nu=1/2$ and 3/2, which is explained by CFs \cite{Jain.2007}. The presence of such a series is unclear in the $N=1$ LL \cite{Pan.PRB.2008}. 

Failure to detect CFs via GR features near $\nu=5/2$ in the $N=1$ LL implies that their mean-free-path (\textit{mfp}) is too small. As mentioned earlier, $n_{CF}^{*}=n/5$ at $\nu=5/2$, assuming fully spin-polarized CFs \cite{footnote_A,Pan.PRL.2011,Pan.PRB.2014,Zhang.PRL.2010,Stern.PRL.2012,Tiemann.Science.2012}. This shrinks the $\nu=5/2$ CFs' $k_F^{*}$ by $1/\sqrt{5}$ compared to the $\nu=1/2$ case because $k_F^{*}=(4\pi n_{CF}^{*})^{1/2}$. We expect the \textit{mfp} to also shrink accordingly. The effect of a small \textit{mfp} on GR features is evident in the comparison between Figs. \ref{fig:Fig1}(e-g). While the $\nu=1/2$ CFs ($n_{CF}^{*}=n$) show even the higher-order ($i=2$ and 3) GR features, we only observe the $i=1$ GR feature for the $\nu=3/2$ CFs ($n_{CF}^{*}=n/3$). It is therefore tempting to argue that $n_{CF}^{*}=n/5$ results in a \textit{mfp} too small to show any GR effect. 
However, the clear GR features observed in Fig. \ref{fig:Fig1}(g), when $\nu=5/2$ resides in the $N=0$ LL ($W\geq50$ nm), provide a strong counterargument. Since the spin-polarized CF density for $\nu=5/2$ is the same in both the $N=1$ and 0 LLs, their \textit{mfp}s should also be comparable in the simplest picture. Yet, no GR features are observed in the $N=1$ LL, unlike the $N=0$ LL. This difference is also manifested in our examination of the GR features near $\nu=7/2$ (Fig. \ref{fig:Fig1}(h)). We observe no GR features when $\nu=7/2$ is in the $N=1$ ($S1$$\downarrow$) LL as for $W=30$ nm (see Fig. \ref{fig:Fig1}(c)). In contrast, when $\nu=7/2$ moves to the $N=0$ ($A0$$\downarrow$) LL for $W=65$ nm, there are clear GR features; here $n_{CF}^{*}=n/7$ is even smaller by a factor of 7/5 than at $\nu=5/2$. This strongly suggests that the \textit{mfp} of CFs in the $N=1$ LL must be anomalously small.



It is possible that in the $N=1$ LL CFs exist only at $\nu=5/2$ and its immediate vicinity but not further away because the ground-state near $\nu=5/2$ is in fact a reentrant integer QHS \cite{Deng.PRL.2012}. 
We emphasize, however, that we do not see any signs of a CF Fermi gas even at $\nu=5/2$ or its immediate vicinity. Note that at all the half-fillings in the $N=0$ LL, there is always a deep and relatively sharp resistance minimum in samples with a one-dimensional density modulation (see, e.g., traces in Figs. \ref{fig:Fig1}(e) and (f)), indicating a ``positive magnetoresistance" as the effective magnetic field for CFs deviates from zero. This is a well-understood Fermi gas property which, similar to the GR features, also stems from the ballistic transport of carriers (with open orbits) under density modulation at very low magnetic fields; it was first seen and explained in 2DESs near zero $B_{\perp}$ \cite{Beton.PRB.1990}, and was later established for CFs near $\nu=1/2$ \cite{Smet.PRL.1999,Zwerschke.PRL.1999}. The absence of positive magnetoresistance in the $N=1$ LL strengthens our previous argument that the LL character may be hindering the ballistic transport of CFs. 

\begin{figure}
\includegraphics[width=.48\textwidth]{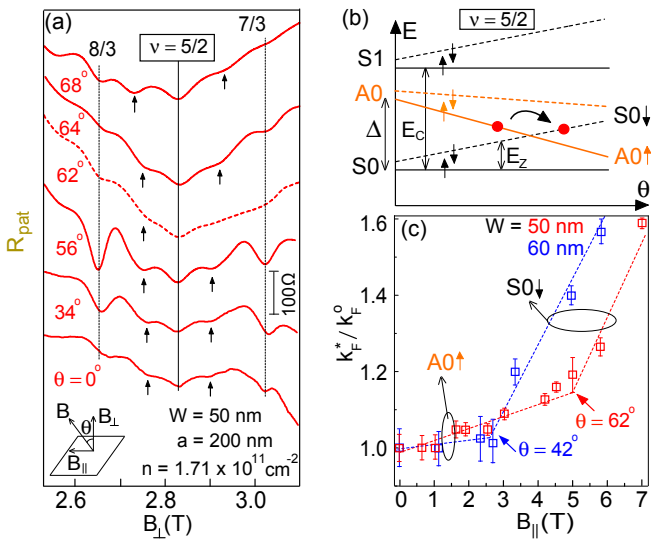}
\caption{\label{fig:Fig2} (color online) (a) Series of magnetoresistance traces near $\nu=5/2$ as a function of $\theta$ for the 2DES with $W=50$ nm and $n=1.71\times10^{11}$ cm$^{-2}$; $\theta$ is the tilt-angle between the field direction and the normal to the 2DES plane (see inset). The thin dotted lines mark the $\nu=8/3$ and 7/3 FQHSs. (b) Schematic LL diagram for the symmetric ($S$) and antisymmetric ($A$) subbands as a function of $\theta$ for the 2DES of (a). 
Solid and dashed lines correspond to the spin-up ($\uparrow$) and spin-down ($\downarrow$) levels. 
(c) Plot of CF Fermi wave vector $k_{F}^{*}$, normalized to its value at $B_{||}=0$, as a function of $B_{||}$ for $W=50$ and 60 nm 2DESs.}
\end{figure}

Having observed CFs at $\nu=5/2$ in the $N=0$ LL of the $A$ subband (for $W\geq50$ nm), we now study them in the $S$ subband's $N=0$ LL. To this end, we tune $\nu=5/2$ from the $A0$$\uparrow$ to $S0$$\downarrow$ LL via applying an in-plane magnetic field ($B_{||}$). Figure \ref{fig:Fig2}(a) inset shows how $B_{||}$ is applied to the 2DES by tilting the sample in field. At tilt angle $\theta=0^{\circ}$, $\nu=5/2$ starts out in the $A0$$\uparrow$ LL (see Fig. \ref{fig:Fig2}(b)) since $E_C>\Delta>$ $E_Z$. As we increase $\theta$, an extra in-plane component ($B_{||}$) is added to the fixed $B_\perp$ at $\nu=5/2$ increasing the spin-splitting $E_{Z}\propto (B_{||}^2+B_{\perp}^2)^{1/2}$ while $E_{C}$ stays fixed at its $\nu=5/2$ value. In contrast, $\Delta$ becomes smaller as $B_{||}$ couples to the 2DES through its finite layer thickness, rendering the charge distribution progressively more bilayer-like and thus reducing $\Delta$ \cite{Lay.PRB.1997,Mueed1.PRL.2015,Hasdemir.PRB.2015}. When $E_Z$ exceeds $\Delta$, $S0$$\downarrow$ crosses with $A0$$\uparrow$ and $\nu=5/2$ moves to the $S0$$\downarrow$ LL (Fig. \ref{fig:Fig2}(b)). 

Signature of the aforementioned LL-crossing is seen in Fig. \ref{fig:Fig2}(a) which presents a series of magnetoresistance traces near $\nu=5/2$, measured for the $W=50$ nm sample at different $\theta$. As $\theta$ increases, the $\nu=8/3$ and 7/3 FQHSs along with the GR features (vertical arrows) initially get stronger for reasons not yet understood. However, all these features weaken significantly at $\theta=62^{\circ}$ (dashed red trace), past which they become stronger again. This weakening of the FQHSs near $\nu=5/2$ is a well-documented signature of LL-crossing \cite{Liu.PRB.2015}. We conclude that the $\nu=5/2$ CFs move from the $A0$$\uparrow$ to the $S0$$\downarrow$ LL in the course of their crossing (Fig. \ref{fig:Fig2}(b)). The GR features near $\nu=5/2$, which yield $k_F^{*}$, thus allow us to quantify their Fermi contour properties in these LLs. In Fig. \ref{fig:Fig2}(c), we plot $k_F^{*}$ (extracted from the stronger GR features to the left of $\nu=5/2$), normalized to the $B_{||}=0$ value ($k_F^{o}$), as a function of $B_{||}$. The increase in $k_F^{*}$ with $B_{||}$ is in agreement with the $B_{||}$-induced elongation of the CF Fermi contour \cite{Kamburov.PRL.2013,Kamburov.PRB.2014}. However, the $k_F^{*}/k_F^{o}$ increase clearly accelerates after $B_{||}\simeq5$ T, or equivalently $\theta\simeq62^{\circ}$, at which the LL crossing takes place. This suggests that the $\nu=5/2$ CFs' Fermi contour is more anisotropic in the $S0$$\downarrow$ than in the $A0$$\uparrow$ LL. Qualitatively similar results are obtained for the $W=60$ nm sample except that the acceleration in $k_{F}^{*}/k_F^{o}$ occurs at a lower $B_{||}$ than the $W=50$ nm case. This is expected: the smaller $\Delta$ at $B_{||}=0$ for $W=60$ nm should induce the LL crossing at a smaller $B_{||}$.

To discuss the different Fermi contour anisotropy of CFs in the $A0$$\uparrow$ and $S0$$\downarrow$ LLs, we consider the Coulomb interaction in the respective LLs. The electron wave function's out-of-plane node in the $A0$$\uparrow$ LL is expected to soften the short range interaction, unlike in the $S0$$\downarrow$ LL. This weaker interaction should cause smaller FQHS energy gaps \cite{Shayegan.PRL.1990} and therefore result in larger CF effective mass \cite{Manoharan.PRL.1994}. According to Ref. \cite{Kamburov1.PRB.2014}, which describes the anisotropy as inversely proportional to the CF effective mass for a given $W$, our observation of smaller anisotropy in the $A0$$\uparrow$ LL indeed points to a larger CF mass, confirming the weaker interaction in the $A$ subband. 

We close by commenting that our data appear to contradict Refs. \cite{Willett.RPP.2013,Willett.PRL.2002} which report ballistic transport of CFs at $\nu=5/2$ in the $N=1$ LL, based on the enhanced conductivity observed in SAW measurements from ultra-high mobility ($\sim2.8\times10^{7}$ cm$^{2}/$Vs) 2DESs. It is possible that the comparatively lower mobility ($\simeq 1\times10^{7}$ cm$^{2}/$Vs) in our 2DESs leads to a CF \textit{mfp} too small for ballistic transport in the $N=1$ LL. While we cannot rule out this possibility, we would like to reiterate an important point. In Refs. \cite{Willett.RPP.2013,Willett.PRL.2002}, the weaker enhanced conductivity observed near $\nu=5/2$ compared to $\nu=3/2$ was attributed to the smaller CF ballistic \textit{mfp} because of the smaller density of CFs near $\nu=5/2$. Our results, however, imply that it is not simply the density of CFs that matters. Note that in our samples, which all have similar mobility and density, we observe clear ballistic transport signatures (positive magnetoresistance and GR features) near $\nu=5/2$ only when the CFs are in an $N=0$ LL. In contrast, we do not see any such features when $\nu=5/2$ lies in an $N=1$ LL, even though the CF density is the same. Our data then imply that it is the LL index that makes the crucial difference and prevents us from observing CF features near $\nu=5/2$ in the $N=1$ LL (assuming that CFs do exist in this LL). In the $N=1$ LL, CFs ought to be more interacting if they are to pair up to form a FQHS at low temperatures. Such interaction, absent in the $N=0$ LL, could act as an extra scattering source for CFs in the $N=1$ LL leading to a smaller $\textit{mfp}$.

\begin{acknowledgments}
We acknowledge support through the NSF (Grants DMR 1305691 and ECCS 1508925) for measurements. We also acknowledge the NSF (Grant MRSEC DMR 1420541), the DOE BES (Grant DE-FG02-00-ER45841), the Gordon and Betty Moore Foundation (Grant GBMF4420), and the Keck Foundation for sample fabrication and characterization. Our measurements were partly performed at the National High Magnetic Field Laboratory (NHMFL), which is supported by the NSF Cooperative Agreement DMR 1157490, by the State of Florida, and the DOE. We thank R. Winkler for providing the charge distribution results in Fig. \ref{fig:Fig1}(a) and J. K. Jain for illuminating discussions. We also thank S. Hannahs, T. Murphy, A. Suslov, J. Park and G. Jones at NHMFL for technical support.
\end{acknowledgments}


\begin{thebibliography}{99}
\bibitem{footnote1}Throughout this paper, we use `` * " to denote the CF parameters.
\bibitem{footnote1_2} The $W=65$ nm sample, whose charge distribution is close to a bilayer as illustrated in Fig. \ref{fig:Fig1}(a), shows weak GR features near $\nu=3/2$ and 1/2 (not shown in Fig. \ref {fig:Fig1}(e)). The interlayer correlation may have weakened the CF state in both cases, leading to weak GR features.
\end{thebibliography}
\end{document}